# Detection of a drag force in G2's orbit: Measuring the density of the accretion flow onto Sgr A* at 1000 Schwarzschild radii

S. Gillessen[1], P.M. Plewa[1], F. Widmann[1], S. von Fellenberg[1], M. Schartmann[1,2], M. Habibi[1], A. Jimenez Rosales[1], M. Bauböck[1], J. Dexter[1], F. Gao[1], I. Waisberg[1], F. Eisenhauer[1], O. Pfuhl[1], T. Ott[1], A. Burkert[1,2], P.T. de Zeeuw[1,3], R. Genzel[1,4]

[1] Max Planck Institute for extraterrestrial Physics, Giessenbachstraße 1, 85748 Garching, Germany
[2] Universitäts-Sternwarte München, Scheinerstraße 1, D-81679 München, Germany
[3] Sterrewacht Leiden, Leiden University, Postbus 9513, 2300 RA Leiden, The Netherlands
[4] Departments of Physics and Astronomy, Le Conte Hall, University of California, Berkeley, CA 94720, USA

*Abstract*
The Galactic Center black hole Sgr A* is the archetypical example of an underfed massive black hole. The extremely low accretion rate can be understood in radiatively inefficient accretion flow models. Testing those models has proven to be difficult due to the lack of suitable probes. Radio and submm polarization measurements constrain the flow very close to the event horizon. X-ray observations resolving the Bondi radius yield an estimate roughly four orders of magnitude further out. Here, we present a new, indirect measurement of the accretion flow density at intermediate radii. We use the dynamics of the gas cloud G2 to probe the ambient density. We detect the presence of a drag force slowing down G2 with a statistical significance of $\approx 9\,\sigma$. This probes the accretion flow density at around 1000 Schwarzschild radii and yields a number density of $\approx 4 \times 10^3$ cm$^{-3}$. Self-similar accretion models where the density follows a power law radial profile between the inner zone and the Bondi radius have predicted similar values.

## *1. Introduction*

The Galactic Center harbors the closest massive black hole (Genzel et al. 2010). Its mass and distance have been determined with ever-higher accuracy since the discovery of stellar orbits around it (Schödel et al. 2002, Ghez et al. 2003, 2008, Gillessen et al. 2009, 2017, GRAVITY collaboration 2017, 2018). The radiative counterpart Sgr A* was discovered by Balick & Brown (1974) at cm-wavelengths. The spectral energy distribution (SED) of Sgr A* rises towards shorter wavelengths and peaks at around 0.3 mm with a flux $\approx 4$ Jy, corresponding to a luminosity of $\approx 50$ L$_\odot$ (Falcke et al. 1998, Brinkerink et al. 2015, Bower et al. 2015, Liu et al. 2016). For the well-determined mass of $4.1 \times 10^6$ M$_\odot$ this corresponds to an extremely low Eddington ratio of $10^{-8}$. Sgr A* is a very underluminous source, despite the gas rich environment in which it resides (Lo & Claussen 1983, Cuadra et al. 2006).

Significant effort went into theoretical studies trying to understand that behavior (Agol 2000, Quataert & Gruzinov 2000a, Falcke & Markoff 2000, Yuan et al. 2003, Pang et al. 2011, Li et al. 2013, Ressler et al. 2018). In the currently favored models of a radiatively inefficient accretion flow (RIAF, see Yuan & Narayan 2014 for a review), the low accretion rate is explained by a combination of a low gas inflow rate at the Bondi radius ($10^5$ R$_S$, Schwarzschild radii), an even smaller accretion rate onto the black hole, and/or a very low radiative efficiency. The models predict a density profile of the gas around Sgr A*, which falls off as n(r) ~ $r^{-\gamma}$ with γ = 1/2 .. 3/2 (Narayan & Yi 1995, Blandford & Begelman 1999, Quataert & Gruzinov 2000b).

Observationally, the gas density has been constrained directly at small and large radius: From measurements of the rotation measure RM one can estimate the inner accretion rate at r ≈ 10 – 100 $R_S$ of $10^{-9}$ – $10^{-7}$ $M_\odot$/yr (Bower et al. 2003, Marrone et al. 2006, 2007). At much larger radii r ≈ $10^5$ $R_S$ X-ray observations with Chandra show a (marginally) resolved X-ray source, whose luminosity and temperature imply a Bondi mass accretion rate of ≈ $10^{-5}$ $M_\odot$/yr (Baganoff et al. 2003). The two constraints bracket four orders of magnitude both in radius and in density, and hence a density profile with γ ≈ 1 is reasonable. Xu et al. (2006) propose

$$n(r) = 1.0 \times 10^4 \; \frac{1}{\text{cm}^3} \left(\frac{r}{10^{15}\text{cm}}\right)^{-1}, \tag{1}$$

where *r* is the distance to Sgr A*. Obtaining further observational constraints on the density profile is difficult due to the lack of suitable probes. In the regime r ≈ $10^3$ – $10^5$ $R_S$ the stellar orbits are observed, but the stars are not affected by the thin and hot accretion flow. This is part of the reason why the discovery of the G2 object (Gillessen et al. 2012) in the S-stars regime attracted attention by theorists (Burkert et al. 2012, Murray-Clay & Loeb 2012, Schartmann et al. 2012, Anninos et al. 2012, Ballone et al. 2013, Guillochon et al. 2014, De Colle et al. 2014). While the physical nature of G2 is a matter of debate, the observed Brackett–γ line emission undisputedly comes from a cloud of gas that cannot be gravitationally bound to any central source. It thus offers a lower density probe in a so far unexplored radius regime, which might interact observably with the RIAF. Plewa et al. (2017) and Steinberg et al. (2018) have derived upper limits on the accretion flow density from the apparent lack of deviation from a simple ballistic Keplerian orbit. Their limit was a factor of a few smaller than what the Xu et al. 2006 law predicts.

G2 might be a knot in a larger gas streamer, as gas is seen ahead of and behind G2. Ahead of G2 is a cloud, first seen in Clénet et al. (2005) and Ghez et al. (2005) as L-band source, which shows similar physical properties, and which revolves on a similar orbit like G2 but with a pericenter passage in 2001 already. Pfuhl et al. (2015) named this source G1 and showed if it initially was on the same orbit as G2, explaining its motion would require a drag force, which in turn constrains the product of cloud cross section and accretion flow density. Assuming the sizes measured for G2 and G1 – with 15 mas ≈ $1.8 \times 10^{15}$ cm these are just below the diffraction limit of current near-infrared instruments at the 8m-class telescopes – remained constant during their pericenter flybys in 2014 (G2) and 2001 (G1), a value of the density close to the Xu et al. (2006) model is implied.

Here, we can strengthen the argument significantly. Now, four years after the pericenter passage of G2, we have a sufficiently long data set to look for a drag force in G2's orbit alone, i.e. without needing to invoke G1. We detect a drag force with a statistical significance of 9σ, and derive a value for the normalization of the density profile, which lies reasonably between the limit given by Plewa et al. (2017) and the Xu et al. (2006) model.

## *2. Observations & Data reduction*

Compared to Plewa et al. (2017) we have been able to add two more years of data. The observations were obtained as in the past with SINFONI at the VLT (Bonnet et al. 2003, Eisenhauer et al. 2003). We used the 25 mas/pix scale and the combined H+K grating (R ≈ 1500) to observe the central arcsecond in the Galactic Center. The prime target of the observations was the star S2 that passed the pericenter of its orbit on May 19, 2018, and for which we have explicitly detected the gravitational redshift (GRAVITY collaboration 2018). Due to the integral field design of SINFONI, these data also contain G2, and we have been able to obtain 870 minutes of high quality on-source integration in 2017 and 1770 minutes in

2018 (table 1). We have used our standard pipeline and wavelength calibration scheme: From the on-source data, we subtracted the sky frames to correct for instrumental and atmospheric background. We applied flat-fielding, bad-pixel correction, a search for cosmic-ray hits, and a correction for the optical distortions of SINFONI. We calibrated the wavelength dimension with line emission lamps and tuned on the atmospheric OH lines of the raw frames. Finally, we assembled the data into cubes with a spatial grid of 12.5 mas pixel$^{-1}$, and created combined data cubes for both years.

| Date | Number of exposures | Date | Number of exposures |
|---|---|---|---|
| 2017-03-14 | 2 | 2018-02-13 | 1 |
| 2017-03-18 | 11 | 2018-02-17 | 4 |
| 2017-04-05 | 10 | 2018-03-23 | 2 |
| 2017-04-16 | 5 | 2018-03-24 | 1 |
| 2017-05-19 | 8 | 2018-03-25 | 12 |
| 2017-05-31 | 4 | 2018-04-08 | 8 |
| 2017-06-01 | 8 | 2018-04-27 | 6 |
| 2017-06-28 | 1 | 2018-04-29 | 7 |
| 2017-07-19 | 7 | 2018-05-03 | 17 |
| 2017-07-27 | 1 | 2018-05-19 | 8 |
| 2017-07-28 | 7 | 2018-05-27 | 8 |
| 2017-07-31 | 1 | 2018-05-29 | 7 |
| 2017-08-18 | 6 | 2018-06-02 | 8 |
| 2017-09-14 | 8 | 2018-06-06 | 10 |
| 2017-09-27 | 2 | 2018-06-22 | 12 |
| 2017-10-17 | 4 | 2018-06-24 | 12 |
| 2017-10-23 | 2 | 2018-07-02 | 14 |
| | | 2018-07-08 | 4 |
| | | 2018-07-27 | 3 |
| | | 2018-08-02 | 3 |
| | | 2018-08-05 | 6 |
| | | 2018-08-18 | 12 |
| | | 2018-08-19 | 12 |

*Table 1: Overview of the SINFONI data collected in 2017 and 2018 used for the analysis here. For each night the evening date of the night is given together with the number of exposures. The numbers are after a quality cut (FWHM of a point source < 7.5 pix in 2017 and < 7.0 pix in 2018). The instrument setup used is the combined H+K-band grating (R ≈ 1500) and the adaptive optics scale yielding a pixel size of 12.5mas. The exposures are single detector integrations of 600 seconds.*

G2 is well detected in both combined cubes, and its appearance is rather compact again, as expected due to the tidal focusing after pericenter passage (Sari et al. 2010). Unlike in the years around pericenter (2013, 2014 and 2015), we can describe its emission by a single position and a single velocity. It is resolved, however, since we detect a velocity gradient (consistent with the orbital direction) across the source.

## 3. Analysis

### 3.1. Position-Velocity-Diagrams

In figure 1 we continue the series of position-velocity diagrams presented in our previous works (Gillessen et al. 2012, 2013, 2013a, Pfuhl et al. 2015, Plewa et al. 2017). These are extracted from the SINFONI cubes along the curved astrometric trajectory of G2, and we show a weighted co-add of the recombination line emission in Brackett-γ (2.166 μm rest wavelength) and He-I (2.058 μm). The slit width was four pixels, and the background is subtracted locally. Further, the diagrams are cleaned by subtracting the respective median values in the horizontal direction (removing artifacts due to the imperfect spectral flattening) and in the vertical direction (removing continuum emission due to stars). Noisy regions (at the edges and around the rest wavelength) are masked out. While the diagrams assume an orbit for G2 in the extraction procedure, we note that the differences between the diagrams when considering different possible orbits for G2 are negligible, since the diffraction limit of the telescope and the pixel size matching the latter are much larger than these differences.

On the red-shifted side, one can see how the initially rather compact object gets tidally stretched ever more along its orbit in the years up to 2012. In 2013, 2014 and 2015 we see the gas both on the red- and blue-shifted side, corresponding to an elongated filament moving around Sgr A* along G2's orbit. From 2016 on, the emission is entirely on the blue-shifted side and continues to move roughly along the orbit. One can see already here that the object appears less blue-shifted than what the Keplerian orbit predicts. This was also the key signature in Pfuhl et al. (2015) for the presence of a drag force, when analyzing motions of G1 and G2 jointly. These diagrams show the tidal evolution of G2, consistent with the motion of an otherwise unbound gas cloud in the gravitational field of Sgr A*. They show that the description as a point mass is not valid in the years around pericenter, and they show the key signature for a drag force, which we will explore in the following.

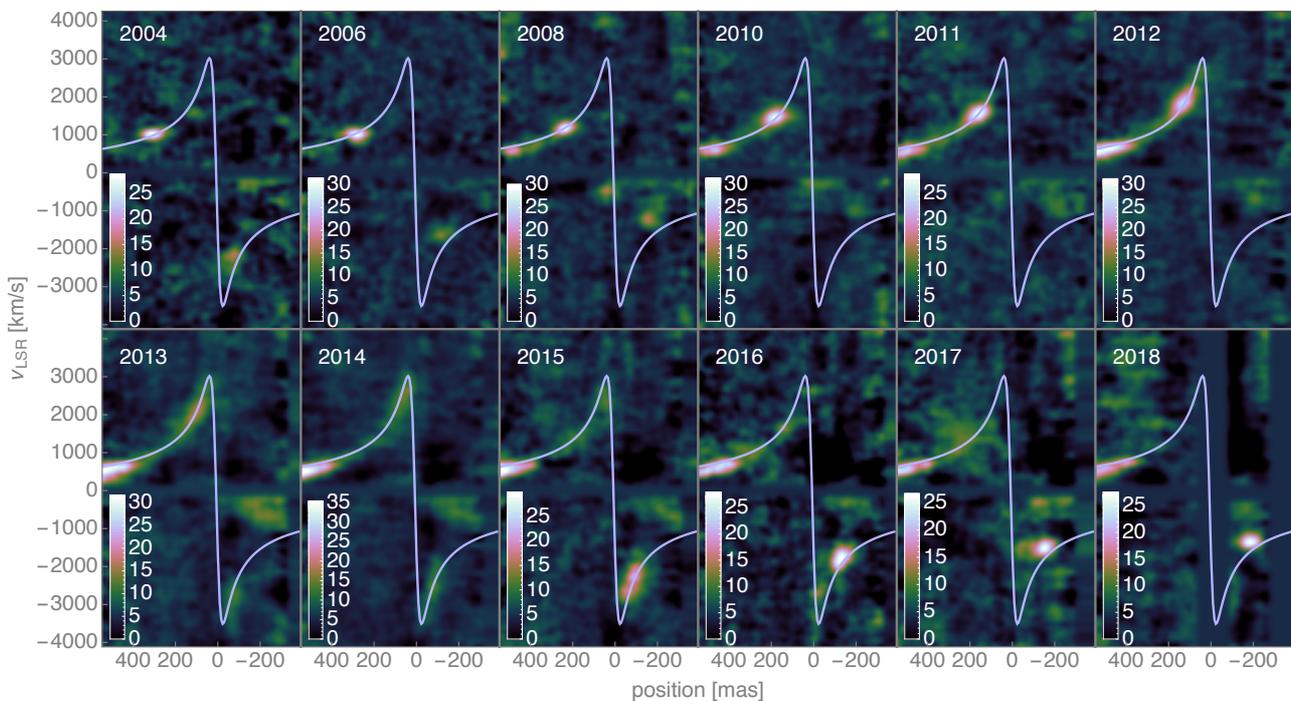

*Figure 1: Series of position-velocity diagrams of G2 extracted from our yearly-combined SINFONI cubes. One can see how G2 initially was redshifted and on approaching Sgr A* got redshifted ever more until 2012. In 2013 - 2015, the gas swings to the blue shifted side and since 2016 G2 appears rather compact again, slowing down on its elliptical orbit moving away from Sgr A*. The blue line is a Keplerian orbit fit to the G2 positional and radial velocity data. The flux units are arbitrary and the scaling is adjusted in each map individually to optimally show the structure of the gaseous emission.*

In the first analysis, we use only the centroid of G2 (in position and velocity) to determine its orbit, like in Pfuhl et al. (2015). In the second analysis, we repeat the more advanced technique presented in Plewa et al. (2017), where a test particle model is fit directly to the 3D data. The former cannot use the data from the years around pericenter, the latter can. Both techniques lead to the same conclusion.

## 3.2. Orbit fits

We extract positions and radial velocities of G2 from the data cubes by means of simple Gaussian fits to the line maps and spectra respectively. The positions thus are from the gas emission, and not from the L-band source (Phifer et al. 2013). The pixel positions are converted to astrometric positions using the astrometry of the unconfused S-stars in the vicinity and a linear transformation (Gillessen et al. 2017). The radial velocities are LSR-corrected. In 2011 and 2012 when G2 was moving quickly, we did not use the yearly cubes but shorter combinations. Overall our data set consists of 13 astrometric positions between 2004 and 2018 and 17 radial velocities between 2003 and 2018. No data from 2013, 2014 or 2015 are used.

For the orbit fits, we assume a fixed potential, which we take from our latest determination in GRAVITY collaboration (2018). We determine thus for a Keplerian fit six parameters (the classical orbital elements), and for a drag force fit we have an additional parameter, encoding the strength of the drag force, following the analysis in Pfuhl et al. (2015). Using a simple ram pressure model for a compressible gas cloud ($C_s \approx 1$) of radius *s* the drag force is (Murray & Lin 2004):

$$F_{\text{drag}} = \frac{\pi}{2} C_s s^2 \rho(r) v^2 \ , \qquad (2)$$

where $\rho(r)$ is the density through which the cloud is plowing, and *v* is the relative velocity between ambient medium and cloud. In the following, we use a few simplifying assumptions:

- We assume that the ambient medium is stationary, i.e. *v* is identical to the orbital velocity. This can only be approximately true, since the accretion flow of Sgr A* has inflow and outflow components.
- The ratio of cross section to mass remains constant. This appears to be a crude assumption, but magnetically arrested models (Shcherbakov 2014, McCourt et al. 2015) could stabilize the cloud geometry.
- The ambient (number) density follows a power law $n(r) = n_0 \times (r/r_0)^{-\gamma}$.

This allows rewriting the drag force in a form suitable for a test particle treatment, where the drag force is proportional to the mass of the object:

$$F_{\text{drag}} = c_D \, r^{-\gamma} \, v^2 \, M_{\text{G2}} \ . \qquad (3)$$

Here, $M_{\text{G2}}$ is the mass of the cloud. The seventh fit parameter is $c_D$, corresponding to the strength of the drag force and encoding the normalization of the density profile $n_0$. If one assumes a density profile with a power law index of $\gamma = 1$, $c_D$ is dimensionless. This is our default assumption.

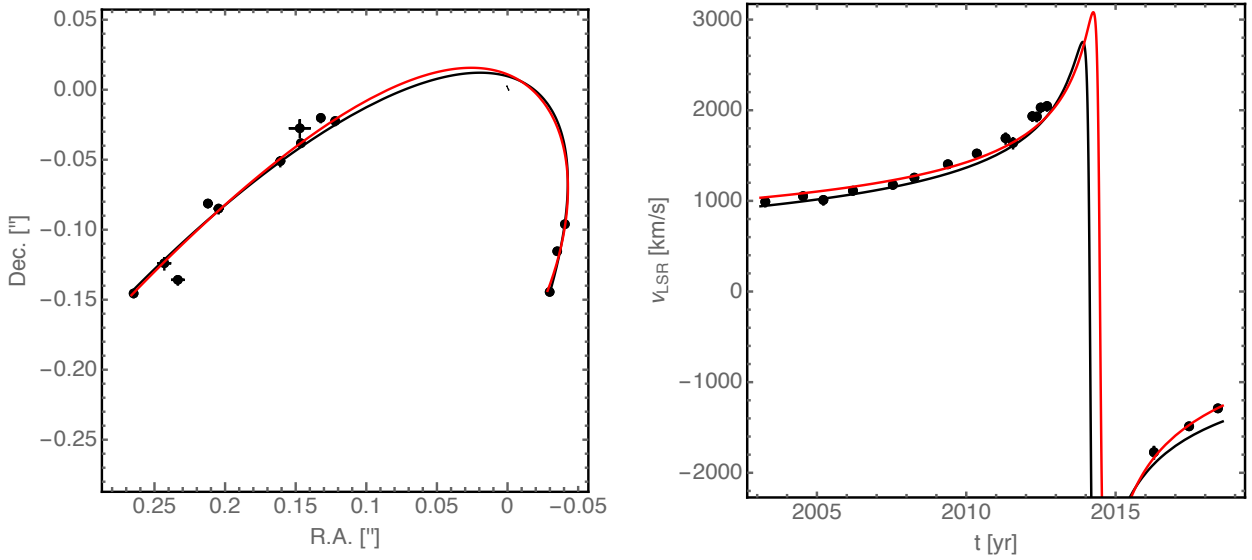

*Figure 2: Orbit fits to the G2 astrometry and radial velocity data. The black data are our SINFONI measurements. The black curves correspond to the best-fitting Keplerian orbit. It does not yield a good fit for the radial velocities post-pericenter (2016 - 2018). The red curves are a drag force orbit, which provides a much better fit to the data. The drag force parameter is statistically significant at the 10 σ level.*

In figure 2 we show orbit fits to the data. The black curves are a Keplerian fit, clearly failing to describe the post-pericenter radial velocity data of G2. One should note that the Keplerian fit is a true best fit in the sense that there does not exist any Keplerian fit that would describe the data better. In red is shown a drag force fit, which describes the data much better. Both fits have a $\chi^2$, which is too large for the degrees of freedom for the fit: 355 / 37 d.o.f. for the Keplerian fit versus 187 / 36 d.o.f. for the drag force fit. The excessive $\chi^2$, however, is caused to a large extent to the pre-pericenter astrometry ($\chi^2$ contribution of 162 for ten 2D data points for the Kepler fit, 101 for the drag force fit). This is due to the difficulty of measuring the G2 position in the earlier cubes: The signal-to-noise ratio is low, the number of reference sources is small, and the integral field unit is not well suited for astrometric measurements.

Nevertheless, the drag force model significantly improves the fit result and $\chi^2$ value compared to the Keplerian model. The best fit yields a drag force parameter of

$$c_D = (3.2 \pm 0.3) \times 10^{-3} \ . \tag{4}$$

Here, the error is the statistical fit error. This is thus a 10 σ significant measurement of the presence of a drag force. In figure 3 we show the parameter correlations obtained from a Markov chain. The drag force parameter $c_D$ is well constrained.

As a null test we have also attempted fitting stellar orbits with the drag force model. For stars with well-determined orbits such as S1 (Gillessen et al. 2017), one obtains essentially the same $\chi^2$ as for the Keplerian fit - however one has used one parameter more, such that the goodness of fit has decreased. The fitted value for the drag force parameter $c_D$ is in these cases consistent with 0, i.e. the error bar exceeds the small numerical value found.

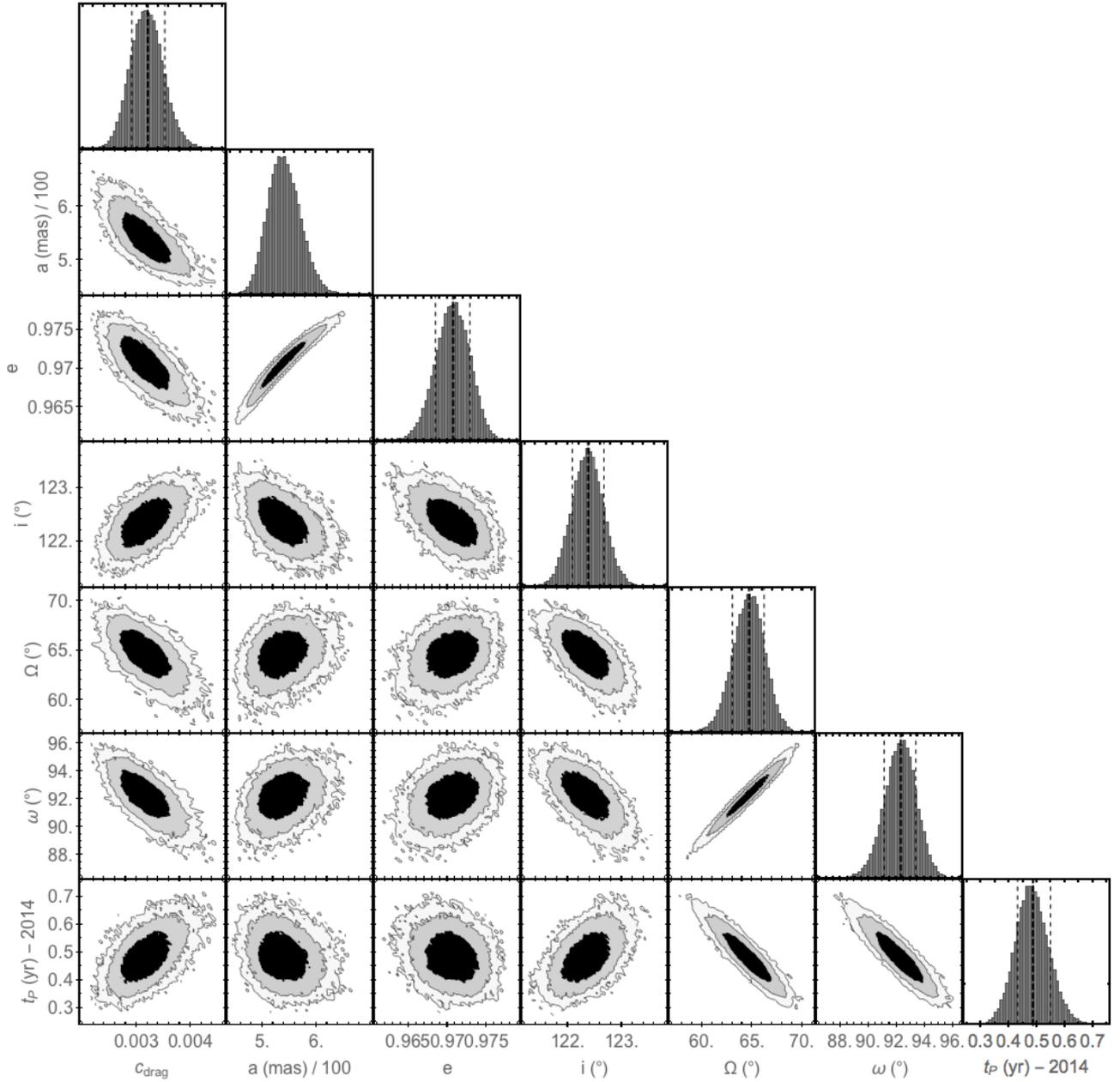

*Figure 3: Posterior probability density distribution of our fit parameters for the drag force fit from a Markov chain. All parameters are well determined. The specific form of the degeneracy between semi-major axis a and eccentricity e means that the parameter that actually is best constrained is the pericenter distance a × (1 − e). The parameter i is the inclination, Ω and ω are the longitude of the ascending node and the argument of pericenter respectively. The parameter $t_P$ is the epoch of pericenter passage.*

A more elaborated treatment is the analysis presented in Plewa et al. (2017). The cloud is modeled as an ensemble of test particles, which independently and without internal interaction move on orbits around Sgr A*. For each set of particles, the corresponding cloud evolves then in time, and it can be fitted at the epochs of observation to the data cubes. In such a scheme not only the orbital elements and the environmental density can be fit parameters, but also parameters describing the cloud, like its size.

We find again, that a Kepler fit does not describe the data well, while a fit with a simple drag force yields a more satisfactory description of both the positional and the radial velocity data (figures 4 and 5). The best fit has a cloud radius $s$ = 22 ± 2 mas at the start of the integration at t = 1971.5 ± 8, and yields for the drag force coefficient

$$c_D = (1.7 \pm 0.2) \times 10^{-3} \ . \tag{5}$$

The error here is again the statistical fit error only, with a significance of around 9 σ. Note that the value obtained is similar to the simpler point mass only model. The two estimates do not agree in the formal, statistical sense, but we have not taken into account any systematic errors introduced by the model assumptions here.

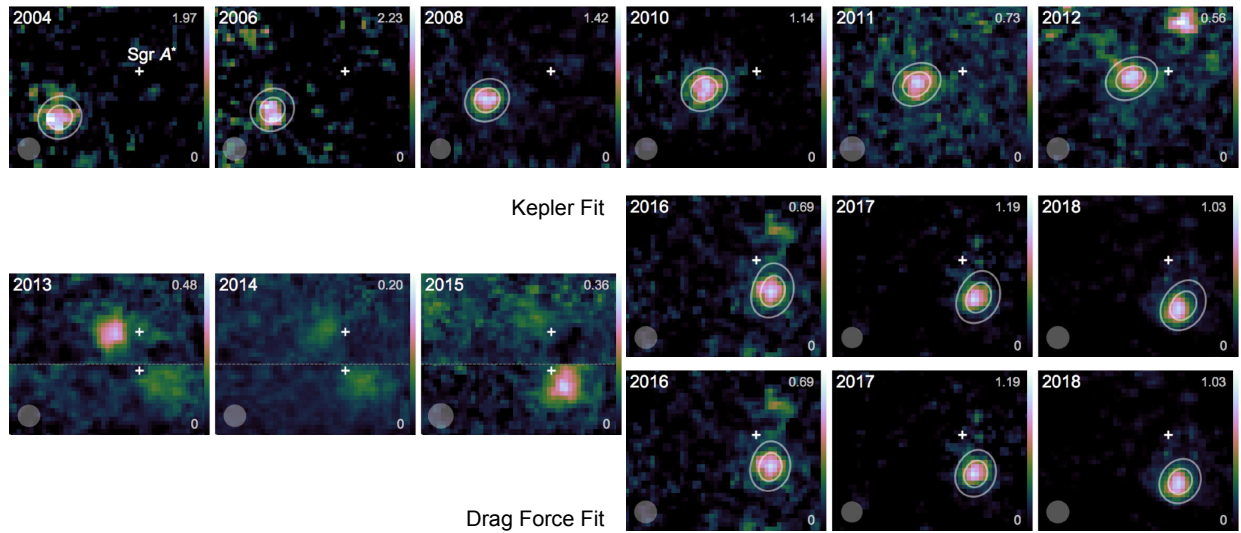

*Figure 4: Brackett-γ channel maps of G2 extracted from our SINFONI data. The panels are normalized to their peak. One can see how G2 approached Sgr A\* (2004 - 2012), then swung by the black hole (2013 - 2015) and now recedes from it (2016 - 2018). The contours give the cloud model fitted to the underlying 3D data. In the years until 2015 the Keplerian and the drag force version of the model are indistinguishable, only in the later years the two models differ visibly. The gray disks denote the point spread function size.*

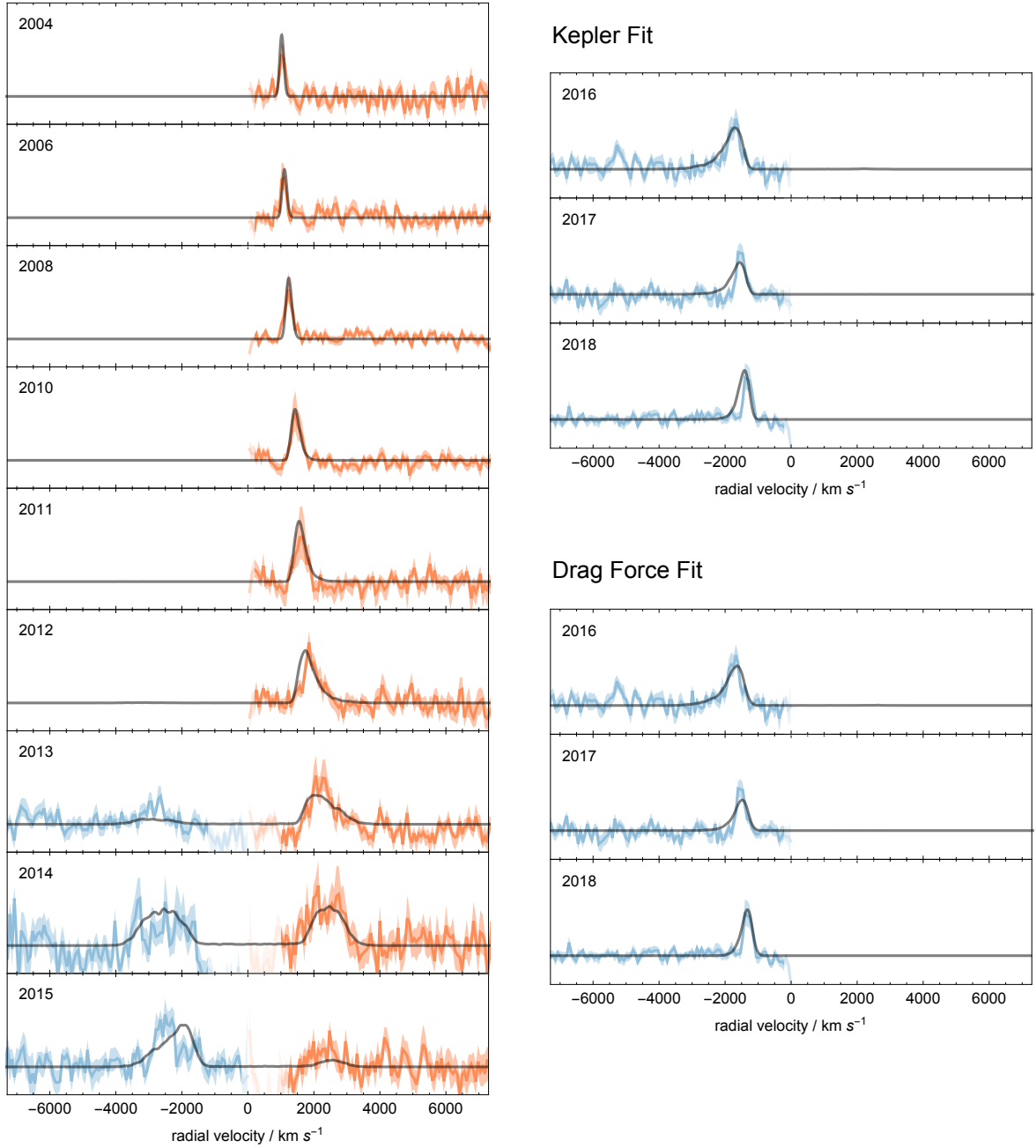

*Figure 5: Spectra of G2 around the Brackett-γ line extracted from our SINFONI data. From 2004 to 2012 G2 got more and more red-shifted, and the line width increased. In 2013 - 2015, the gas swung over to the blue-shifted side, corresponding to the pericenter passage of G2. Since 2016 G2 decelerates, as it moves away from the black hole. The solid lines show the cloud model fitted to the underlying 3D data. The differences between the Kepler and the drag force model are noticeable only in the last three years. In particular, the 2017 and 2018 panels show that the drag force model is a better fit to the data.*

### 3.3 Conversion to normalization of the density profile

The mass of (the gaseous component of) G2 is reported in Gillessen et al. (2012) to be

$$M_{G2} = 3\sqrt{f_V}\left(\frac{s}{15\text{ mas}}\right)^{3/2} M_{\text{Earth}}, \qquad (6)$$

with $f_V \approx 1$ the volume filling factor of G2. Equating the two expressions for $F_{\text{drag}}$ (equations 1 and 2) yields for the normalization of the number density

$$n_0 = 0.25 \, c_D(\gamma) \frac{f_V^{\frac{1}{2}}}{c_s \, s^{\frac{1}{2}} \, r_0^{\frac{5}{2}}} \frac{M_{\text{Earth}}}{m_p} \left(\frac{r_0}{\text{mpc}}\right)^{-\gamma}, \tag{7}$$

where $r_0 = 10^{15}$ cm is the normalization radius and $m_p$ the proton mass. The canonical, measured value for $s$ is $s = 15$ mas $\approx 1.82 \times 10^{15}$ cm (Gillessen et al. 2012, Pfuhl et al. 2015). Using this value the simple point mass fit yields

$$n_0 = (6.6 \pm 0.7) \times 10^3 \, \frac{1}{\text{cm}^3}, \tag{8}$$

where the error is still the statistical error only. For the cloud fit with $s = 22$ mas we obtain

$$n_0 = (3.5 \pm 0.4) \times 10^3 \, \frac{1}{\text{cm}^3}. \tag{9}$$

While the detection of a drag force is significant at the 8 – 10 σ level, the conversion of the drag force parameter into a density comes with systematic uncertainties.

Our measurement depends on the assumed source size $s$, since $n_0 \sim s^{-1/2}$ (equation 7). Assuming a source size between 10 mas and 30 mas thus adds an error of $\approx 32\%$ for $n_0$.

The density profile slope $\gamma$ is another source of uncertainty. For Bondi accretion (Bondi 1952) $\gamma = 3/2$. Using that profile we obtain $c_D = 1.1 \times 10^{-3}$ and $n_0 = 4.0 \times 10^3$ cm$^{-3}$. On the other hand, the recent Chandra data (Shcherbakov & Baganoff 2010, Wang et al. 2013) suggest a flatter profile for the outer part around $10^5$ R$_S$, with a slope around $\gamma = 1/2$, for which we find $c_D = 8.3 \times 10^{-3}$ and $n_0 = 9.7 \times 10^3$ cm$^{-3}$. The uncertainty on the profile introduces thus an error of order $\approx 40\%$ on $n_0$.

Further, we assumed the volume filling factor $f_V$ of the cloud to be unity. Another approximation is the cloud shape, which if different from spherical would also change the drag force. The errors introduced by the uncertainty of the gravitational potential used for the orbit fit are negligible.

Overall, we estimate that our measurement is good to a factor of $\approx 4$:

$$n_0 = (2..8) \times 10^3 \, \frac{1}{\text{cm}^3} \tag{10}$$

## Discussion

Our measurement of the G2 orbit is the first dynamical measurement of a hydrodynamic drag force outside of the solar system. As a consequence of the drag, G2 should not rush back out to where the apocenter of the orbit previously was (probably at around 70 mpc $\approx 1.8$" in the years around 1850), but the energy loss makes the orbit shrink significantly. In our simple point mass model, G2 should revert its motion inwards already in 2045, reaching a maximum distance from Sgr A* of 23 mpc $\approx 0.6$'' only. The inspiral would end around 2155 - but obviously the approximation of the cloud by a point mass breaks down much before.

Comparing with the G1-G2-based estimate (Pfuhl et al. 2015) shows that our G2-only estimate yields a drag force that is weaker by a small factor. In Pfuhl et al. (2015) we found $c_D = (8.4 \pm 1.3) \times 10^{-3}$. The difference is not surprising, given the additional assumptions needed for G1's and G2's orbits. Also note that the G2-only drag force estimate is the smaller of the two. That would be consistent with G1 plunging through the medium first, and clearing the path for G2 falling in a few years later (Pfuhl et al. 2015), but we note that replenishing of a hole with radius $s = 15$ mas $\approx 2 \times 10^{15}$ cm should happen on a time scale τ of $\tau \approx s \, / \, c_{\text{sound}} \approx 2 \times 10^7$ s < 1 year, much shorter than the temporal distance between G1 and G2.

The limits derived in Plewa et al. (2017) correspond to $n_0 = (1.9 \pm 0.7) \times 10^3$ cm$^{-3}$. Hence, our density estimate is moderately larger than the limit derived there. This might be an effect due to the data around pericenter, which are not particularly well described in neither of our models, see figure 5. In Plewa et al. (2017) the weight of these data are relatively higher, potentially biasing the fit. Our value range, however, seem less compatible with the simulations in Steinberg et al. (2018), who set a limit of $n_0 \approx 10^3$ cm$^{-3}$. These simulations have been the only ones capable of reproducing the near-to-constant Brackett-γ flux of G2. For higher densities, the cloud would fragment due to cooling instabilities and the Brackett-γ flux would have increased by a large factor.

Our measured density is comparable, albeit probably a bit lower than what the model of Xu et al. (2006) predicts. In figure 6 we compare our measurements to the other measurements of the accretion flow density and the Xu et al. (2006) model. Our density estimate nicely fills in the gap at intermediate radii around $r \approx 10^3$ $R_S$, and matches overall remarkably well, given that very different ways of trying to assess the density of the hot gas around Sgr A* are employed.

Our data are not very sensitive to the slope γ of the ambient gas density profile. Attempting to fit for it yields $\gamma = 1.9 \pm 1.3$. The reason is the combination of high eccentricity and the correspondingly sharp peak of the velocity around pericenter, the ram pressure model having a force $\sim v^2$ and the radial profile peaking inwards $\sim r^{-1}$, which means that the energy loss almost completely happens around pericenter. In effect, we are not able to distinguish between a continuous slow-down and an instantaneous kick around pericenter.

We adopted a simple 1- and N-particle model in order to investigate the effect of a drag force on G2. This approach is motivated by its remarkable (and perhaps surprising) success in reproducing the kinematical and geometrical evolution of G2 (Gillessen et al. 2012, 2013ab, 2017, Pfuhl et al. 15, Plewa et al. 2017), perhaps understandable in a picture in which G2 consists of individual droplets, i.e. $f_V < 1$. Yet, treating such a system of gas clumplets as independent particles is unrealistic, since the downstream part of G2 will be shielded by clumplets at the front.

Our models neglect much of the physics expected to be relevant for the (magneto-) hydrodynamic evolution of G2 (Schartmann et al. 2012, 2015; Sadowski et al. 2013, Shcherbakov 2014). The interaction of the two gaseous media will lead to local effects, such as deformations, shocks or hydrodynamic instabilities. Further, the external gas pressure will be important, as might be internal magnetic pressure. It is worth noting that the theoretical models (for example in Schartmann et al. 2015) did not predict an overall slow-down of G2.

McCourt et al. (2015) show that for a magnetized G2 model the magnetic field of the accretion flow will enhance the drag force, giving in principle access to the Alfven speed of the ambient medium. The effect, however, is degenerate with the ambient medium density. Including magnetic fields thus will lead to lower estimates for the density, since part of the drag force will be of magnetic nature. In this light our value might overestimate the density.

McCourt & Madigan (2016) used the G1/G2 data set to show that a drag force model including a rotating accretion flow is a better description of the data than the simple model with a static ambient medium. The improvement came mostly from a better description of G1's radial velocities, and requires a few additional parameters describing the orientation and the speed of the accretion flow. In principle, the same treatment can also be applied to the G2 data set used in this work. However, we note that unlike for the G1/G2 data set, we do not see an obvious, systematic deviation from our simple model in the data. It is thus less justifiable to add more parameters to our model at this point. Yet, future orbital data of G2 might require such an extension of the model, and the work by McCourt et al. provides a framework for this.

Madigan et al. (2017) have noted that if the ambient medium around Sgr A* is inflowing, the resulting radial forces would lead to a prograde precession of the orbital ellipse, effectively delaying the pericenter passage by a few months compared to the Keplerian prediction. This is a natural way to explain the (small) differences in the orbital elements between G1 and G2 (Witzel et al. 2017). Interestingly, this picture is also consistent with what we see from our G2-only fits: The drag force orbit has the closest approach at 2014.58 ± 0.13. This needs to be compared to the the Keplerian orbit as known pre-pericenter, i.e. using data up to and including 2012. The pericenter date for that fit is 2014.30 ± 0.13, which means the significance of that number being different from the one in the drag force fit, is below 2σ, Δt = 0.28 ± 0.18 years. Yet the sign and size of the effect (if real) remarkably match the model in Madigan et al. (2017). Similarly, the fit also determines by how much the orbital ellipse has rotated. We find a prograde shift of 2.4° ± 1.9°. With future data we can hope to improve the uncertainties in the orbit measurements. Given current error bars, we might reach the regime in which we can constrain the radial motion of the accretion flow.

Our measurement is in direct conflict with the conclusion of Witzel et al. (2014), who have claimed that G2 needs to be stellar in nature. These authors have claimed to rule out the gas cloud scenario based on the observed compactness of G2 after pericenter in L-band (3.8μm) observations. Given the well-resolved tidal evolution seen over 14 years in our position-velocity diagrams (figure 1), their conclusion seems to be premature. Given the observed size of the gas in G2, an implausibly large central mass would be needed to bind the gas against the tidal forces of the massive black hole. It is thus unavoidable that G2's gas is tidally evolving - and that it can be used as a probe for the ambient medium. Yet, the compactness of the L-band source might be compatible with the gas cloud model. The L-band source is continuum emission from ≈ 600K warm dust. One could thus think of a source consisting of a hidden central star with a compact, unresolved dust envelope and an unbound gas shell. In this picture, the central object with its dust shell would continue to move on the original Kepler orbit, while the gas has lost kinetic energy during the pericenter passage, now following a different orbital trace. The difference between the two orbits can only be traced astrometrically (since the L-band source does not show any spectral features) and amounts to ≈ 2 mas in 2018, which is comparable to the uncertainties by which one can determine the position of the L-band source and the gas emission. However, the difference will increase in the coming years, reaching around 5 mas in mid 2019 and more than 10 mas by 2021. This will allow testing the compound scenario of a gas cloud around a dust-enshrouded star.

Another, completely different explanation for a detecting a deviation from the original Kepler orbit was proposed by Zajacek et al. (2014): G2 might have initially been a binary, and we were lucky enough to catch the tidal break up of the system at pericenter passage (Hills 1988). One would need to assume that the second partner has escaped unnoted, which would mean that it is a low-mass star or a dark remnant. Further, this model does not address the tidal evolution of the gas emission.

Overall and given the uncertainties, we have a robust detection of a drag force in G2's orbit, yielding a measurement of the accretion flow density at an intermediate radius. The agreement of our measurement with theoretical and previous estimates is satisfactory, and given previous theoretical work, one might prefer a value close to the lower end of our regime. This would suggest that in the innermost region of the accretion flow the particle density is around $10^6$ cm$^{-3}$, which for standard parameters of the accretion flow would mean that the plasma is highly magnetized.

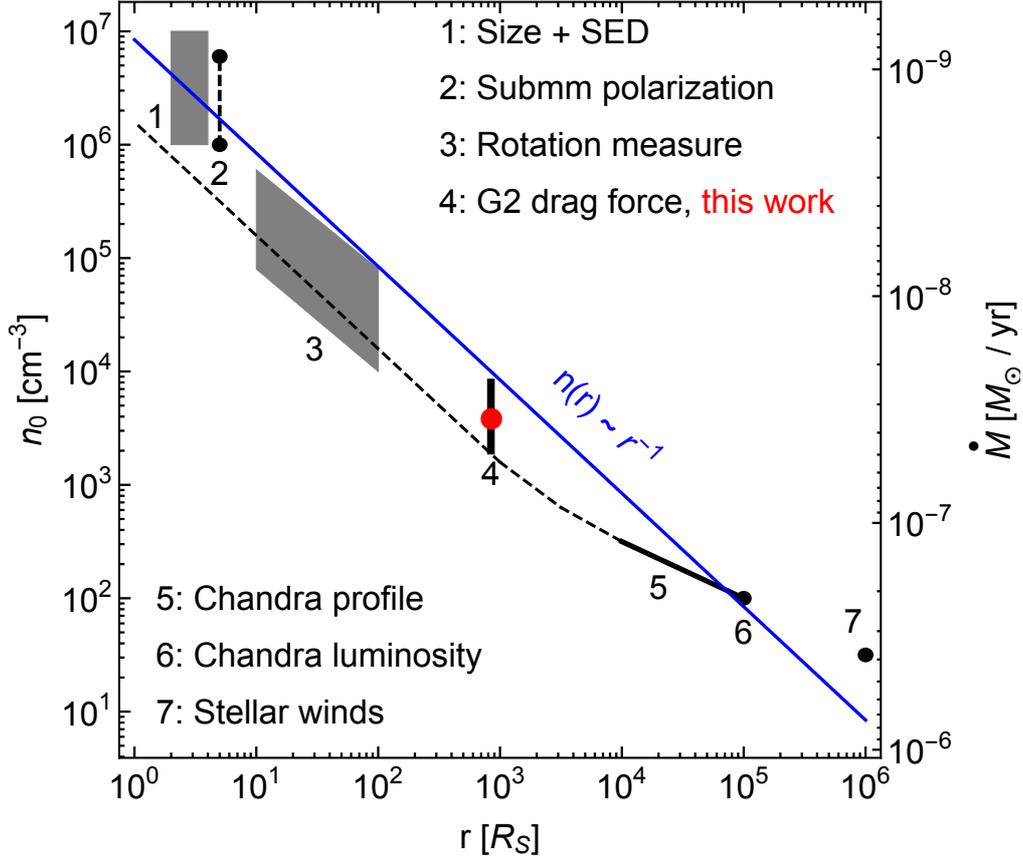

*Figure 6: The measured density profile (left ordinate axis) of Sgr A*'s accretion flow. Measurement 1 comes from one zone models of the inner accretion zone using the resolved size (Doeleman et al. 2008) and flux at the submm peak of the SED (Bower et al. 2015, von Fellenberg et al. 2018). These are consistent with the results of detailed models using relativistic MHD simulations (e.g., Moscibrodzka et al. 2009, Dexter et al. 2010, Shcherbakov et al. 2012). Measurement 2 uses the submm polarization and flux (Agol 2000, Quataert & Gruzinov 2000a). Measurement 3 is based on the rotation measure (Bower et al. 2003, Marrone et al. 2006, 2007). This work is measurement 4. The X-ray profile as measured in Wang et al. (2013) from Chandra data is shown as measurement 5, normalized to match the value in Baganoff et al. (2003, measurement 6). Measurement 7 is the value derived in Quataert et al. (2004), who simulate the accretion onto Sgr A* from stellar winds. The blue line is the model from Xu et al. (2006). The black , dashed line tries to reconcile the measurements at all radii, predicting a density of $10^6$ cm$^{-3}$ in the central region. The right ordinate axis shows the corresponding accretion rate for standard accretion theory assuming a scale height H/R = 0.3, viscosity parameter α = 0.1, and a virial ion temperature.*